\documentclass[prl,superscriptaddress,twocolumn]{revtex4-1}
\usepackage[latin9]{inputenc}
\usepackage{color}
\usepackage{amsmath}
\usepackage{amssymb}
\usepackage{graphicx}
\usepackage{verbatim}

\usepackage[unicode=true,
bookmarks=true,bookmarksnumbered=false,bookmarksopen=false,
breaklinks=true,pdfborder={0 0 1},backref=true,colorlinks=true]
{hyperref}
\hypersetup{
	linkcolor=magenta, urlcolor=blue, citecolor=blue, pdfstartview={FitH}, hyperfootnotes=true, unicode=true}

\makeatletter

\usepackage{amsfonts}\usepackage{tabularx}\usepackage{dcolumn}\usepackage{bm}\usepackage{graphicx}\usepackage{epstopdf}
\usepackage{times}
\hypersetup{urlcolor=blue}

\makeatother

\begin{document}
	
	\title{Solving Quantum Master Equations with Deep Quantum Neural Networks}

	\author{Zidu Liu}
	\affiliation{Center for Quantum Information, IIIS, Tsinghua University, Beijing 100084, P. R. China}
	\author{L.-M. Duan}\email{lmduan@tsinghua.edu.cn}
	\affiliation{Center for Quantum Information, IIIS, Tsinghua University, Beijing 100084, P. R. China}
	\author{Dong-Ling Deng}\email{dldeng@tsinghua.edu.cn}
	\affiliation{Center for Quantum Information, IIIS, Tsinghua University, Beijing 100084, P. R. China}
	\affiliation{Shanghai Qi Zhi Institute, 41th Floor, AI Tower, No. 701 Yunjin Road, Xuhui District, Shanghai 200232, China}
	\date{\today}

\begin{abstract}
Deep quantum neural networks may provide a promising way to achieve quantum learning advantage with noisy intermediate scale quantum devices. Here, we use deep quantum feedforward neural networks capable of universal quantum computation to represent the mixed states for open quantum many-body systems and introduce a variational method with quantum derivatives to solve the master equation for dynamics and stationary states. Owning to the special structure of the quantum networks, this approach enjoys a number of notable features, including the absence of barren plateaus, efficient quantum analogue of the backpropagation algorithm, resource-saving reuse of hidden qubits,  general applicability independent of dimensionality and entanglement properties, as well as the convenient implementation of symmetries. As proof-of-principle demonstrations, we apply this approach to both one-dimensional transverse field Ising and two-dimensional $J_1-J_2$ models with dissipation, and show that it can efficiently capture their dynamics and stationary states with a desired accuracy.
\end{abstract}

\maketitle

\begin{figure}[h]
\includegraphics[width=0.4\textwidth]{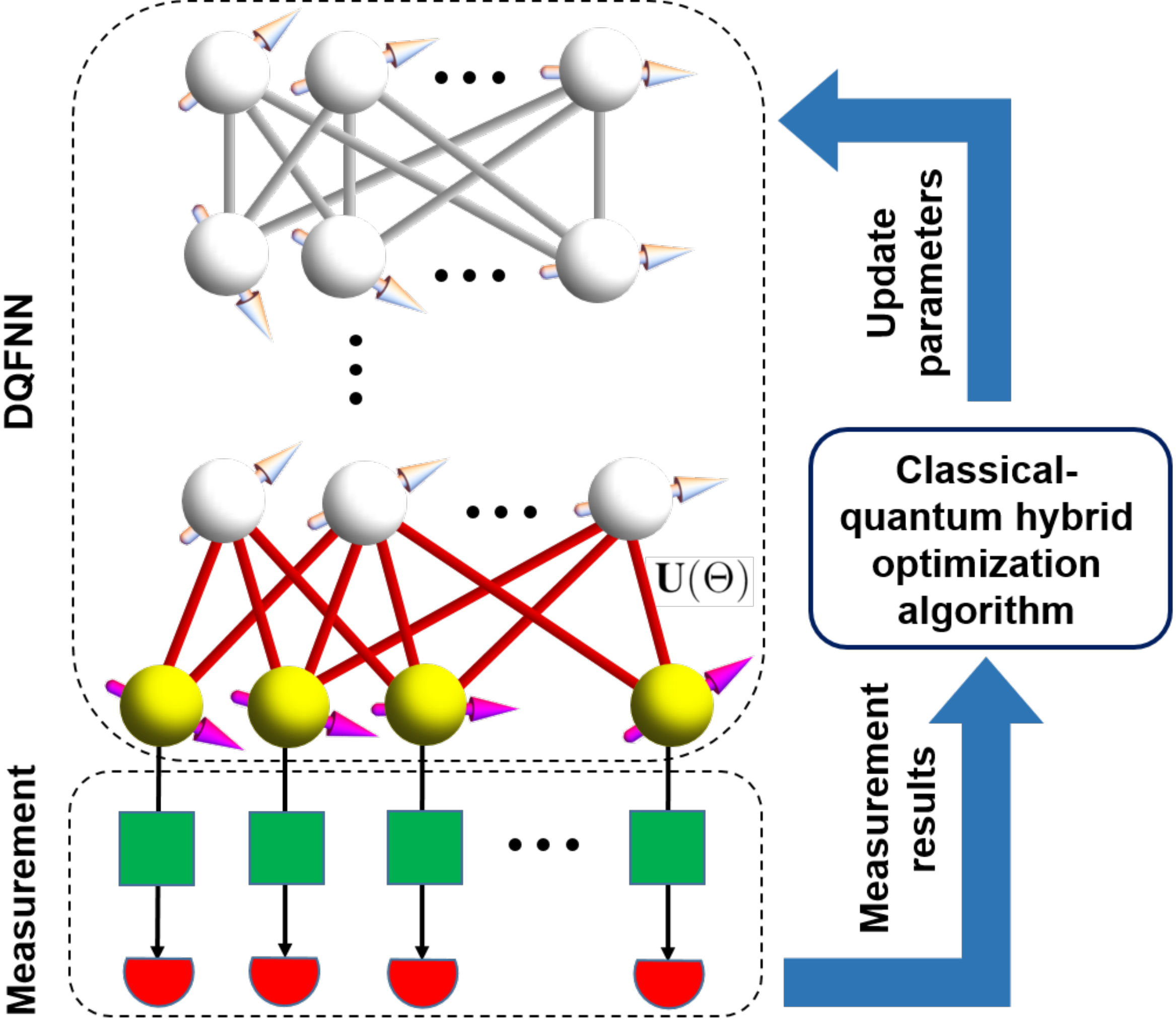}
\caption{Schematic illustration of the DQFNN (deep quantum feedforward neural network) method in solving quantum master equations. This DQFNN has $d$ layers in total, with an input (top), output (bottom), and $d-2$ hidden layers. A quantum perceptron is defined as an arbitrary unitary operator acting on qubits from neighboring layers \cite{beer2020training}, and the DQFNN is parametrized by the product of these unitary operators. We use the output state as the variational ansatz state of a quantum open system. The qubits in the output layer are measured on different basis and the measurement results are used by a classical-quantum hybrid optimization algorithm to update the network parameters.
}
\label{fig:figure5}
\end{figure}

Recent developments in quantum hardware have reached a stage where quantum devices with tens to hundreds of controllable qubits will soon become available  \cite{arute2019quantum,song2019generation,krantz2019quantum,wright2019benchmarking,bruzewicz2019trapped,Levine2019Parallel}. An important milestone for these so-called noisy intermediate-scale quantum (NISQ) devices \cite{Preskill2018quantum} is Google's demonstration of quantum supremacy with $53$ programmable superconducting qubits \cite{arute2019quantum}. This opens up exciting possibilities of utilizing NISQ devices to tackle challenging real-world problems that are beyond the capacity of any classical computers \cite{Harrow2017Quantum}. Along this line, a variety of hybrid quantum-classical approaches have been pushed forward, including variational quantum eigensolver (VQE)  \cite{VQE2014variational,vqekokail2019self,Liu2019Variational,Wang2019Accelerated},  and quantum approximate optimization algorithm (QAOA) \cite{QAOAfarhi2014quantum,Zhou2020Quantum,qaoaMoll_2018}, quantum classifiers \cite{Schuld2020Circuit,havlivcek2019supervised,cong2019quantum}, and quantum generative adversarial networks \cite{Lloyd2018Quantum,Demers2018Quantum,hu2019quantum}, etc.  In this paper, based on deep quantum feedforward neural networks \cite{beer2020training} (DQFNNs) we introduce a variational method with quantum derivatives to obtain the dynamics and stationary states for open quantum many-body systems by solving the Lindblad master equations (see Fig. \ref{fig:figure5} for a pictorial illustration).

Every quantum system is inevitably coupled to its surrounding environment. In most cases, the coupling is of Markovian type and the dynamics of open quantum systems are governed by the Lindblad quantum master equation \cite{breuer2002theory}. Thus, solving this master equation plays a key role in studying open quantum systems. However, for quantum many-body systems this turns out to be a formidable challenge due to the exponential growing of the Hilbert space dimension with the size of the system. To combat this challenge, a number of prominent numerical approaches based on classical computers have been developed, such as these using tensor network representations \cite{Zwolak2004Mixed,Verstraete2004Matrix, Orus2008Infinite,cuiTNS.114.220601,wernertns.116.237201,Gangat2017Steady,jaschkeTns2018one,KshetrimayumTns} or quantum Monte Carlo methods \cite{Yan2018Interacting,Nagy2018Driven}. More recently, machine-learning inspired approaches based on artificial neural networks \cite{carleo2017solving}, especially the restricted Boltzmann machine (RBM), have also been introduced to solve the quantum master equation \cite{CarleoOpenRBM,CiutiOpenRBM,VinceOpenRBM,RyusukeOpenRBM}.
Each of these approaches bears its own advantages and disadvantages, and the choice of which one to use is problem-specific. For instance, tensor-network approaches are typically very effective for one-dimensional (1D) systems involving small entanglement, but may face pronounced difficulties in higher dimensions or in situations where massive entanglement comes into play \cite{schollwock2011density,verstraete2008matrix}. Quantum Monte Carlo methods rely on efficient sampling of physical configurations \cite{Foulkes2001Quantum} and are effective for certain open quantum systems  \cite{Yan2018Interacting,Nagy2018Driven}, yet could suffer from a severe sign problem that is common in simulating dissipative dynamics \cite{hangleiter2019easing}. The neural-network approaches \cite{CarleoOpenRBM,CiutiOpenRBM,VinceOpenRBM,RyusukeOpenRBM}  are generally applicable to high dimensional systems and entanglement is not a limiting factor \cite{Deng2017Quantum}.  However, like in most machine-learning tasks \cite{goodfellow2016deep},  their performance depends crucially on the suitable tuning of hyperparameters. In fact, the capability and limitation of neural-network methods in solving quantum many-body problems still remain largely unclear and related studies are at the research forefront.

Here, inspired by the classical neural-network approaches and the exciting experimental progress in developing NISQ devices, we propose a variational DQFNN method to solve the quantum master equations. We note that variational algorithms running on NISQ devices for open quantum systems have been discussed in the literature \cite{yuan2019theory,endo2018variational,hu2020quantum,yoshioka2019variational}, but to the best of our knowledge most existing works focus on straightforward variational quantum circuits that may suffer from the notorious barren plateau (i.e., vanishing gradient) problem \cite{Mcclean2018Barren,Cerezo2020Cost}. 
We utilize a deep quantum neural network, which is arranged in layers and convenient for the quantum analogue of the backpropagation algorithm, to serve as the ansatz density state of the open quantum system.
We adopt a hybrid quantum-classical stochastic reconfiguration (SR) algorithm \cite{sorella2007weak} to variationally solve the quantum master equation in the Lindblad form to obtain the dynamics and steady state. As a result of the special structure of the quantum networks, our approach escape barren plateaus \cite{beer2020training} and allows for convenient implementation of translational symmetry and resource-saving reuse of hidden qubits. In addition, it works for generic open quantum many-body systems, independent of dimensionality,  the amount of entanglement involved, and the specific forms of interaction and dissipation. We benchmark our approach with both dissipative 1D transverse field Ising and 2D $J_1-J_2$ models. Our results pave a way to explore the rich physics of open quantum systems with quantum neural networks that would be implemented using NISQ devices in the near future.



\textit{The general problem and DQFNN structure.}\textemdash We consider the problem of solving the following master equation: 
\begin{eqnarray}
\label{masterequation1}
    \frac{d\rho}{dt}=\mathcal{L}\rho = -i[H,\rho]-\sum_i\frac{\gamma_i}{2}(\{F^{\dagger}_i F_i,\rho\}-2F_i \rho F^{\dagger}_i), \label{MasterEq}
\end{eqnarray}
%
where $\rho$ is the density matrix of the system, $H$ is the Hamiltonian governing the unitary part of the dynamics, $F_i$ are the jump operators describing the dissipative processes induced by the environment with dissipative rate denoted by $\gamma_i$, the curly bracket represents the anticommutator, and $\mathcal{L}$ is the Liouvillian superoperator  \cite{Heinz2007The}.

%

We use the final output state of a quantum neural network, the recently proposed DQFNN in particular \cite{beer2020training}, to serve as the variational ansatz for the density state of the open quantum system. The network structure is illustrated in  Fig.\ref{fig:figure5}. Each node represents a qubit and a quantum perceptron is an arbitrary unitary operator acting on  several qubits from neighboring layers. Information propagates from the top (input) layer to the bottom (output) layer.  At the $i$-th layer, the output state $\rho^{(i)}_\text{o}$ is determined by
\begin{eqnarray}
\rho_{\text{o}}^{(i)} = \text{Tr}_{i-1}[\textbf{U}(\rho_{\text{o}}^{(i-1)}
    \otimes
    |\psi^{(i)}\rangle \langle \psi^{(i)}|)
    \textbf{U}^\dagger], \label{i-layer output state}
\end{eqnarray}
where $\rho^{(i-1)}_\text{o}$ is the output state of the $(i-1)$-th layer (which serves as the input state for the $i$-th layer), $|\psi^{(i)}\rangle \langle\psi^{(i)}|$ denotes the initial state of the $i$-th layer, $\textbf{U}= \otimes_{j=0}^M U_{i,j}$ represents the total unitary operation between the two layers with $U_{i,j}$ denoting the $j$-th perceptron at the $i$-th layer, and  $\text{Tr}_{i-1}$ denotes a partial trace of the $(i-1)$-th layer. The Eq. (\ref{i-layer output state}) may also be regarded as a mapping $\mathcal {M}$ from $\rho^{(i-1)}_\text{o}$ to $\rho^{(i)}_\text{o}$, and the whole network features a mapping-composition structure, which is essential for the backpropagation algorithm \cite{goodfellow2016deep}:
\begin{eqnarray}
\rho^{(d)}_{\text{o}} = \mathcal{M}^{d}(\cdots\mathcal{M}^{1}( |\psi^{(1)}\rangle \langle\psi^{(1)}|)\cdots).
\end{eqnarray}
Here, $|\psi^{(1)}\rangle \langle\psi^{(1)}|$ denotes the initial state of the first (input) layer and we suppose that the network has $d$ layer in total. We mention that this quantum network is universal in the sense that it can map any input density state to an arbitrary output state due to its vast flexibility in designing the qubit perceptrons. In fact, it has been shown in Ref. \cite{beer2020training} that such a DQFNN can carry out universal quantum computation even for perceptrons as simple as unitaries acting on two-input and one-output qubits. Here, we focus on  two-input one-output qubit perceptrons for simplicity and experimental feasibility.

\textit{The general recipe.}\textemdash As mentioned, we use $\rho^{(d)}_{\text{o}}$ to serve as the variational ansatz state and adapt a hybrid quantum-classical SR algorithm \cite{sorella2007weak} to variationally solve Eq. (\ref{MasterEq}) to obtain the dynamics and steady state of the open quantum system. More specifically, we solve the following optimization problem variationally:
\begin{eqnarray}
\min_{\Theta}\;  ||\partial_{t}\rho^{(d)}_{\text{o}}(\Theta)-\mathcal{L}\rho^{(d)}_{\text{o}}(\Theta)||_{\text{FS}},
\end{eqnarray}
where $\Theta$ collectively denotes all the parameters used to describe the qubit perceptrons (unitary gates) and  the Fubini-Study (FS) norm is used to measure the distance. For convenience, we omit the  labels and rewrite the density matrix $\rho^{(d)}_{\text{o}}$ as a vector $\vec{\rho}$. Similar to the SR algorithm in solving the quantum master equation based on the variational RBM representation \cite{CarleoOpenRBM, VinceOpenRBM}, the above minimization problem results in the following system of equations:
\begin{equation}
    \sum_{\nu} S_{\mu, \nu} \dot{\Theta}_{\nu}=f_{\mu},
\end{equation}
where $\Theta_{\nu}$ denotes the $\nu$-th parameter,  and
	\begin{eqnarray}
	S_{\mu, \nu} &=\text{Re}(\frac{\partial \vec{\rho}^\dagger}{\partial \Theta_\mu}\frac{\partial \vec{\rho}}{\partial \Theta_\nu})-\text{Re}[(\frac{\partial \vec{\rho}^\dagger}{\partial \Theta_\mu}\vec{\rho})(\vec{\rho}^\dagger\frac{\partial \vec{\rho}}{\partial \Theta_\nu})],\\
	f_{\mu} &=\text{Re}(\frac{\partial \vec{\rho}^\dagger}{\partial \Theta_\mu}\mathcal{L}\vec{\rho})-\text{Re}[(\frac{\partial \vec{\rho}^\dagger}{\partial \Theta_\mu}\vec{\rho})(\vec{\rho}^\dagger\mathcal{L}\vec{\rho})].
	\end{eqnarray}
We initialize the parameters to be small random values and update them iteratively according to the the following rule:
\begin{eqnarray}
\Theta^{\text{new}}=\Theta^{\text{odd}}+\lambda S^{-1} \vec{f}, \label{ParameterUpdate}
\end{eqnarray}
where $\lambda$ is a hyperparameter (called the learning rate in the machine learning literature) introduced to modulate the convergence  of the iteration process.

In real experiments, both $S$  and $\vec{f}$ might be obtained from measurements of observables or their linear combinations (see the Supplemental Material \cite{DQFNNsupp} for details). Then, by using Eq. (\ref{ParameterUpdate}) we update the parameters of the quantum neural network iteratively and obtain the dynamics of the open system consequently. At long time limit, the neural network state converges to the steady state of the system. As discussed in Ref. \cite{beer2020training}, a striking feature of this DQFNN approach is that the parameter matrices can also be calculated layer-by-layer and at any given time we need only access to two layers. This will greatly reduce the memory requirements of the algorithm and enables a resource-saving reuse of the qubits.

\begin{figure}
\centering
\includegraphics[scale=0.35]{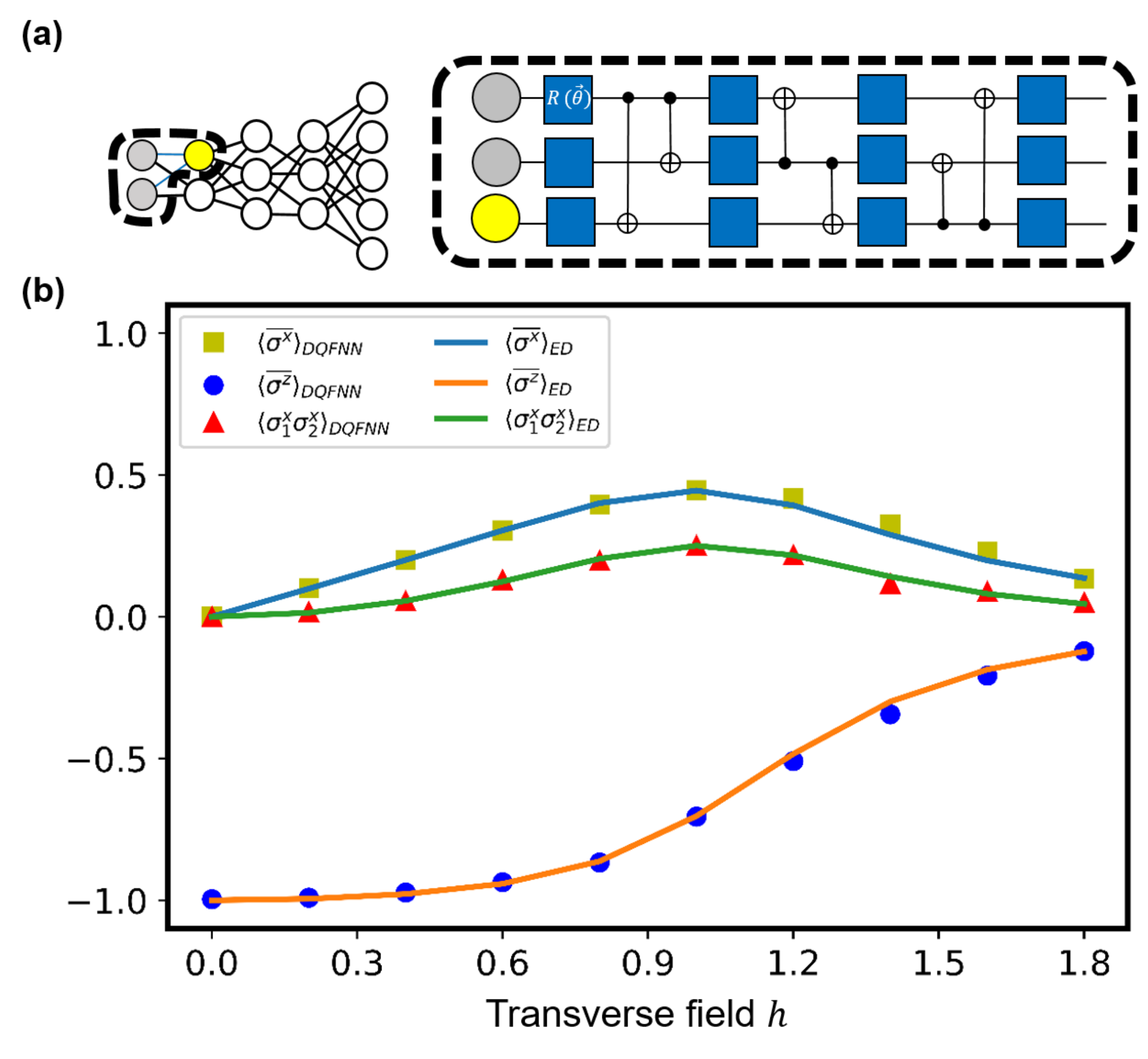}
\caption{ (a) A DQFNN with layer structure $(2,2,3,3,5)$ is used to solve the dissipative transverse field Ising model [see Eq. (\ref{Eq:1DIsingModel})]. Here, each quantum perceptron consists of a unitary acting on two-input and one-output qubits and this unitary is parametrized by a simple quantum circuit (black-dotted box), which contains twelve Euler rotations and six controlled-not gates.
(b)Numerical results for the steady state of the dissipative Ising model with five spins. The expectation values of three different observables $\overline{\sigma^x}=\frac{1}{5}(\sum_k \sigma_k^x)$, $\overline{\sigma^z}=\frac{1}{5}(\sum_k \sigma_k^z)$, and $\sigma^x_{1}\sigma^x_{2}$ are calculated by both the DQFNN and exact diagonalization (ED) method. Here, we set $J=1$ as the energy unit and the dissipation ratio is $\gamma=1$.  For the DQFNN approach, the learning rate $\lambda$ is chosen to be $\lambda=0.01\times0.999^{N_{s}}$, where $N_{s}$ is the number of steps.
}
\label{fig:NNstructure and SSresults}
\end{figure}

\textit{Numerical simulations.}\textemdash We apply the introduced DQFNN approach to two concrete models to benchmark how it works. The first example involves the dissipative transverse field Ising model in one dimension:
\begin{eqnarray}\label{Eq:1DIsingModel}
 \frac{d \rho}{d t} &=& -i [H, \rho]-\frac{\gamma}{2}\sum_j  [\{\sigma_j^+\sigma_j^{-},\rho\}-2\sigma_j^-\rho \sigma_j^+ ],
\end{eqnarray}
where $H = J\sum_{j}\sigma_j^z \sigma_{j+1}^z + h \sum_{j}  \sigma_j^x$, $\gamma$ denotes the dissipation rate, and $\sigma_j^{\pm}=\frac{1}{2}(\sigma_j^x\pm i\sigma_j^y)$, with $\sigma^{x,y,z}_j$ being the Pauli matrices for the site $j$.  The first term in the Hamiltonian represents the $z$-component spin-spin interaction in the longitudinal direction with strength $J$ and the second term denotes a local uniform magnetic field with strength $h$ along the transverse direction. We mention that the steady state of this model has been studied in Ref. \cite{CiutiOpenRBM} with a classical neural-network (i.e., RBM) approach. Here, instead we apply the DQFNN method to obtain both the dynamics and steady state. The quantum neural-network structure is shown in Fig.\ref{fig:NNstructure and SSresults} (a). The system size is $N=5$ and the quantum network has five layers and contains $15$ qubits in total.  Each perceptron consists of three qubits coupled by a series of unitary gates as shown by the quantum circuit in the dotted box in Fig.\ref{fig:NNstructure and SSresults} (a).  We consider the periodic boundary condition and implement the translational symmetry for reducing the number of parameters.  In addition, owing to the special structures of the quantum perceptrons, the derivatives of the density states with respect to the parameter $\Theta_{\nu}$ can be obtained by \cite{Liu.diff.RRA.98.062324}:
\begin{equation}\label{eq: derivative}
    \frac{\partial \rho(\Theta_{\nu})}{\partial \Theta_{\nu}} = \frac{1}{2}[\rho(\Theta_{\nu}+\frac{\pi}{2})-\rho(\Theta_{\nu}-\frac{\pi}{2})].
\end{equation}
We mention that the derivatives in Eq. (\ref{eq: derivative}) is exact, in sharp contrast to the finite-difference method that may inevitably induce a small differential error. This might be important for the convergence of the optimization procedure  \cite{Harrow2019Low}.

\begin{figure}
\includegraphics[width=0.48\textwidth]{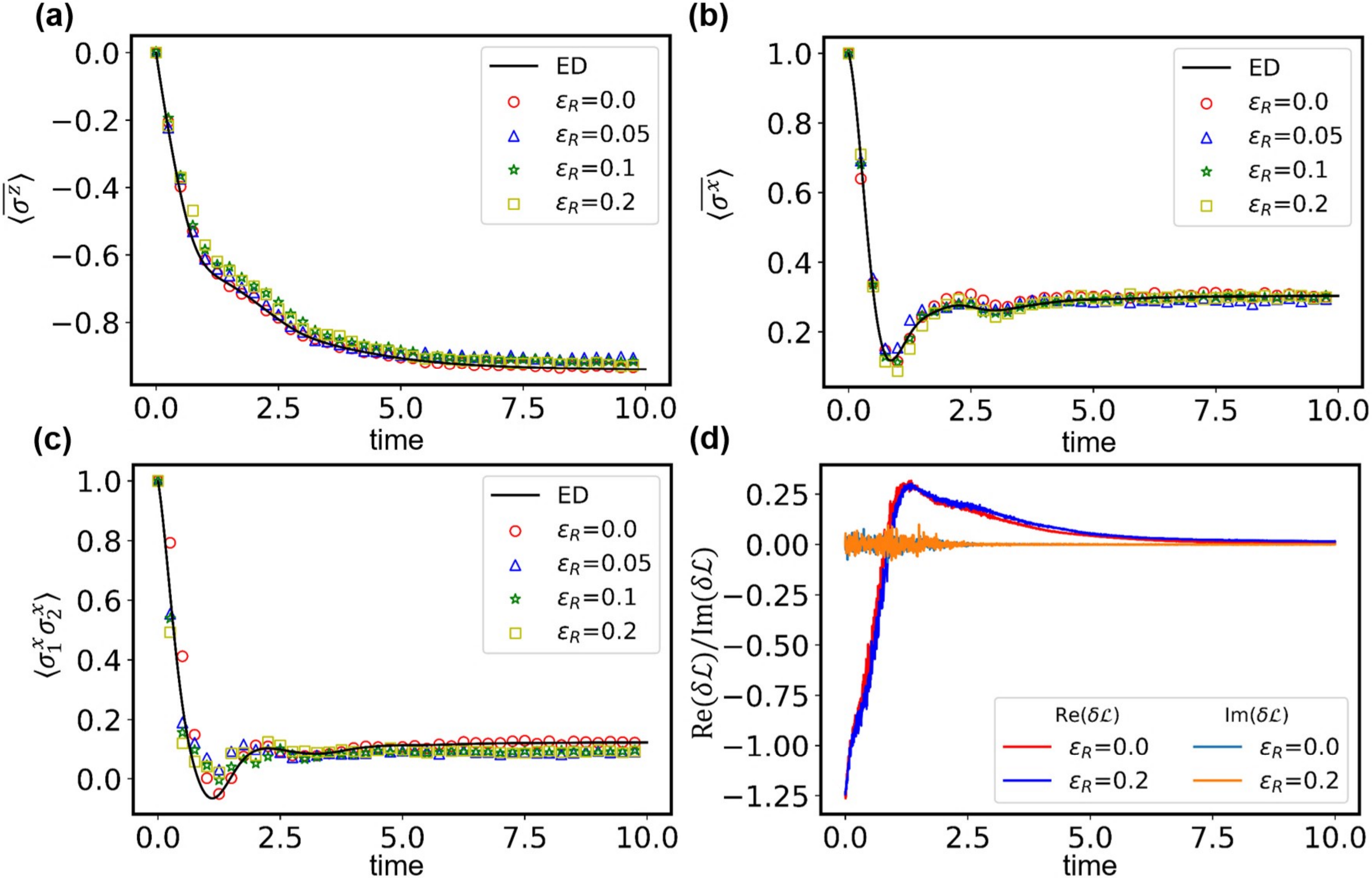}
\caption{Numerical results for the time dynamics of the dissipative Ising model with five spins under the periodic boundary condition. 
For the DQFNN approach, random Gaussian noises with varying strength $\epsilon_R$ are added to the quantum derivatives to account for the imperfections in real experiments.  The time step is chosen to be $5\times 10^{-3}$  and the sampling size for every step is $5\times 10^4$. (a), (b), and (c) plot the time dependences of  $\overline{\langle \sigma^z \rangle}$,  $\overline{\langle \sigma^x\rangle}$, and $\langle \sigma^x_1 \sigma^x_2 \rangle$, respectively. (d) plots the real and imaginary parts of $\delta \mathcal{L}$, which measures the convergence to the steady state. }
\label{fig:IsingDynamics}
\end{figure}

With classical computers, we numerically simulate the quantum neural networks and the whole process of using the DQFNN  approach to obtain the dynamics and steady state for the dissipative Ising model defined in Eq. (\ref{Eq:1DIsingModel}). We start with a DQFNN state where all the constituted qubits are initially set to be $|+\rangle=(|0\rangle+|1\rangle)/\sqrt{2}$ and all the involved parameters are initially set to be a small value near zero. Then the parameters are updated according to the optimization procedures shown in Eq.(\ref{ParameterUpdate}). Partial of our results are shown in Fig. \ref{fig:NNstructure and SSresults}(b) and Fig. \ref{fig:IsingDynamics}. 
In Fig. \ref{fig:NNstructure and SSresults}(b), we compute the steady state with varying strength of the transverse field $h$ through the DQFNN approach.  We plot the magnetization values $\overline{\langle \sigma^{x,z}\rangle}$ and correlations  $\langle \sigma^x_1\sigma^x_2\rangle$ for the stationary state, and compare them with the ED results. It is clear that the DQFNN results match the ED results excellently. 
In Fig. \ref{fig:IsingDynamics}, we consider the dynamics of our DQFNN during the optimization process with fixed $h=0.6$. Here, we add a random Gaussian noise into the quantum derivatives to account for the imperfections in real experiments and to verify the robustness of our DQFNN approach. 
In Fig. \ref{fig:IsingDynamics}(a,b,c), we plot the magnetization values $\overline{\langle \sigma^{x,z}\rangle}$ and correlations $\langle \sigma^x_1\sigma^x_2\rangle$ with different noise strength, and compare their corresponding results obtained by ED. From these figures, it is evident that the noiseless DQFNN results match the ED results (with relative error smaller than $10^{-2}$) and converge to the corresponding values for stationary state at long time. The added noise does cause small deviation during the dynamic process, while the evolution process eventually converges to the steady state as well. In Fig. \ref{fig:IsingDynamics}(d), we show $\delta \mathcal{L}$, which is the expectation value of the Lindblad operator to measure the convergence to the stationary state,  as a function of time. We note that $\delta \mathcal{L}$ approaches zero as we increase the iterations steps, implying that the density state of the system represented by the quantum neural network converges to the stationary state indeed, despite the addition of strong noises. 

%


\begin{figure}
\includegraphics[scale=0.32]{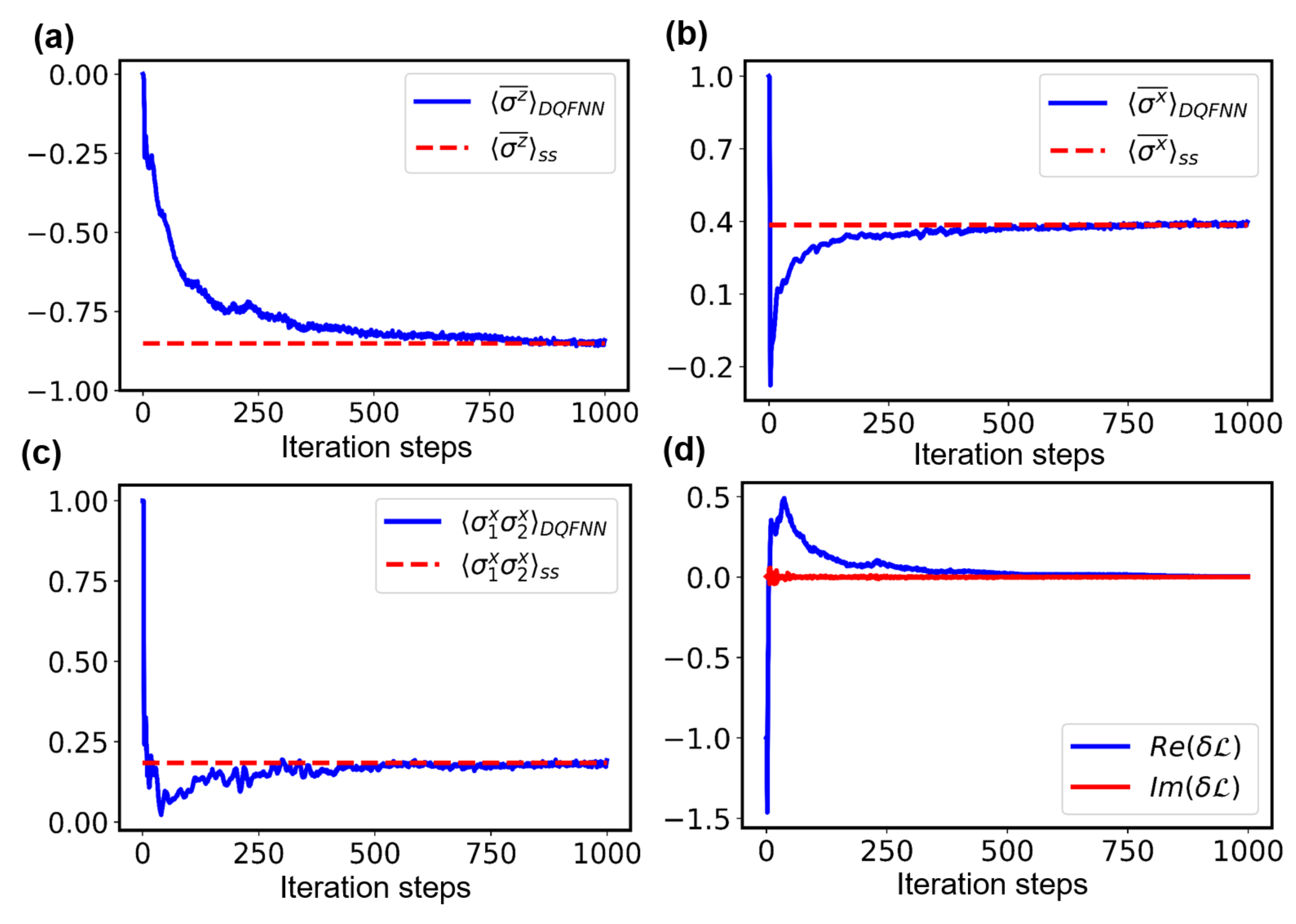}
\caption{Numerical results for the dissipative 2D $J_1-J_2$  model. Here, the simplest two-by-two square lattice is considered and we use a DQFNN with layer structure $(4,4,4)$ to obtain the results. The parametrization of the DQFNN is similar to the case for the 1D dissipative Ising model. (a), (b), and (c) show respectively $\overline{\langle \sigma^z\rangle}$, $\overline{\langle \sigma^x\rangle}$, and $\langle \sigma^x_0\sigma^x_1\rangle$ as a function of iteration steps with the dissipation rate chosen to be $\gamma=1.0$. The red-dotted lines represent their corresponding values for the steady state (SS) obtained from exact diagonalization. (d) plots both the real and imaginary parts of $\delta \mathcal{L}$, which converges to zero after around $1000$ iteration steps ($|\delta \mathcal{L}|<10^{-2}$), indicating that the output state of the DQFNN indeed converges to the steady state.
}
\label{fig:2DJ1J2}
\end{figure}

Our second example concerns the dissipative $J_1-J_2$ model defined on a square-lattice in 2D with Hamiltonian:
\begin{eqnarray}
H = J_1 \sum_{\langle i,j\rangle}\sigma^z_i\sigma^z_{j} +  J_2 \sum_{\langle\langle i,j\rangle\rangle}\sigma^z_i\sigma^z_{j} + h \sum_{i} \sigma^x_i,
\end{eqnarray}
where the first (second) term represents the nearest-neighbor (next-nearest-neighbor) $z$-component spin-spin interaction with strength $J_1$ ($J_2$). The dissipation considered here is the same as that for the 1D Ising case [see Eq. (\ref{Eq:1DIsingModel})]. For this Hamiltonian, it is easy to observe that there is a geometric frustration to the antiferromagnetic state due to the competition between the first and second terms. 
For simplicity, we set $J_1=1$ as the energy unit, $J_2=\frac{1}{2}J_1$, $h=J_1$, and the dissipation rate  $\gamma=J_1$. Our numerically simulated results from the DQFNN approach are shown in Fig. \ref{fig:2DJ1J2}. In Fig. \ref{fig:2DJ1J2}(a,b,c), we plot the magnetization values $\overline{\langle\sigma^{x,z}\rangle}$ and the correlation function $\langle \sigma^x_1\sigma^x_2\rangle$  as a function of iteration steps. As shown, it is clear that all these quantities converges nicely to their corresponding exact values for the steady state (after $1000$ iterations, their relative errors are all smaller than $10^{-2}$). This indicates that after around $1000$ iterations, the output state $\rho^{(d)}_{\text{o}}$ of the DQFNN indeed converges to the steady state, which is also clearly manifested in Fig. \ref{fig:2DJ1J2} as $\delta \mathcal{L}$ converges to zero (after $1000$ iterations, $|\delta \mathcal{L}|$ becomes less than $10^{-2}$).



We remark that the accuracy of our DQFNN results can be improved in a number of ways, including increasing the number of layers of the quantum neural network or the sampling size at each iteration step, tuning hyperparameters, and designing more appropriate quantum perceptrons, etc. Due to limited classical computation resources,  in this paper we only carry out simulations for small systems. However, this will not be a problem if quantum devices are used in practice. As discussed in the Supplementary Material \cite{DQFNNsupp}, the time complexity of the DQFNN approach is about $O(N^3)$, which indicates its scalability to larger systerms. 
We also stress the difference between the RBM \cite{CarleoOpenRBM,CiutiOpenRBM,VinceOpenRBM,RyusukeOpenRBM} and DQFNN approaches: the RBM method is entirely classical and the gradients that are crucial in updating parameters rely on efficient Monte Carlo sampling; in contrast, for the DQFNN method the gradients can be obtained directly from measurements of observables in experiment \cite{DQFNNsupp}. This might lead to an advantage in  computational cost for the DQFNN approach for certain problems where no efficient sampling scheme is available. In the future, it is of  both fundamental and practical interest to carry out a complete study on the advantages and limitations for the DQFNN approach.

\textit{Conclusion.}\textemdash We have introduced a deep quantum neural-network method to solve Lindblad master equations for open quantum many-body systems. Through concrete examples in both 1D and 2D, our results demonstrated that both the dynamics and stationary states of such systems can be efficiently obtained via this method with a desirable accuracy. Due to the special structures of DQFNNs, our approach is generally applicable to high dimensional systems and is independent of the amount of entanglement involved.  In addition, it is robust to experimental imperfections and free from the barren plateau problem, and allows a convenient implementation of translational symmetry and a resource-saving reuse of qubits. This classical-quantum hybrid approach is particular suitable for running on NISQ devices in the near future.

	We thank Yuanhang Zhang, Sirui Lu, and Xun Gao for helpful discussion. This work was supported by the National key Research and Development Program of China (Grant No. 2016YFA0301902), Tsinghua University, and the Ministry of Education of China. DLD also would like to acknowledge additional support from the Shanghai Qi Zhi Institute.

\bibliographystyle{apsrev4-1-title}
\bibliography{DQNNreferences}

\clearpage
\onecolumngrid
\setcounter{figure}{0}
\makeatletter
\renewcommand{\thefigure}{S\@arabic\c@figure}
\setcounter{equation}{0} \makeatletter
\renewcommand \theequation{S\@arabic\c@equation}
\renewcommand \thetable{S\@arabic\c@table}

\begin{center} 
	{\large \bf Supplementary Material for: Solving Quantum Master Equations with Deep Quantum Neural Networks}
\end{center}

	\section{classical-quantum hybrid updating scheme}
	We first introduce a classical-quantum hybrid Monte Carlo sampling method which can approximate the mean value of a operator efficiently. Same as the classical Monte Carlo method, the statistic error decays with the sampling size $N_{s}$ in $\frac{1}{\sqrt{N_{s}}}$\cite{Metropolis1953MC}. Consider a general hermitian operator $\mathcal{A}$, its expectation value can be obtained by sampling  a Monte Carlo chain with the probability $\rho_{l,l}=\langle l|\rho|l\rangle$ \cite{CarleoOpenRBM}: 
	\begin{equation}
	\langle\mathcal{A} \rangle = \sum_{l} \rho_{l,l} \frac{(\mathcal{A}\rho)_{l,l}}{\rho_{l,l}}
	=\sum_{l} \rho_{l,l} \sum_{m} \frac{\mathcal{A}_{l,m} \rho_{l,m}}{\rho_{l,l}}.
	\end{equation}
	Thus, $(\mathcal{A}\rho)_{l,l}$ can be computed efficiently if the variable $\mathcal{A}$ consists of only local operators.
	%
	As for the non-hermitian operators, take the $S$ and $\vec{f}$ as an example, we firstly rewrite the $S$ and $\vec{f}$ in an explicit form:
	\begin{equation}
	\begin{aligned}
	f_{\mu} &= \text{Re}\left[\sum_{\vec{l}, \vec{r}}|\rho_{\vec{l}, \vec{r}}|^2 (\frac{1}{\rho_{\vec{l}, \vec{r}}}\frac{d\rho_{\vec{l}, \vec{r}}}{d\Theta_\mu})^\ast\frac{(\mathcal{L}\rho)_{\vec{l},\vec{r}}}{\rho_{\vec{l},\vec{r}}}-\sum_{\vec{l}, \vec{r}}|\rho_{\vec{l}, \vec{r}}|^2 (\frac{1}{\rho_{\vec{l}, \vec{r}}}\frac{d\rho_{\vec{l}, \vec{r}}}{d\Theta_\mu})^\ast
	\sum_{\vec{l'}, \vec{r'}}|\rho_{\vec{l'}, \vec{r'}}|^2\frac{(\mathcal{L}\rho)_{\vec{l'},\vec{r'}}}{\rho_{\vec{l'},\vec{r'}}}\right],
	\end{aligned}
	\end{equation}
	and
	\begin{equation}
	\begin{aligned}
	S_{\mu, \mu'} &= \text{Re}\left[\sum_{\vec{l},\vec{r}}|\rho_{\vec{l},\vec{r}}|^2
	(\frac{1}{\rho_{\vec{l}, \vec{r}}}\frac{d\rho_{\vec{l}, \vec{r}}}{d\Theta_\mu})^\ast\frac{1}{\rho_{\vec{l}, \vec{r}}}\frac{d\rho_{\vec{l}, \vec{r}}}{d\Theta_\mu'}-\sum_{\vec{l},\vec{r}}|\rho_{\vec{l},\vec{r}}|^2(\frac{1}{\rho_{\vec{l}, \vec{r}}}\frac{d\rho_{\vec{l}, \vec{r}}}{d\Theta_\mu})^\ast
	\sum_{\vec{l'},\vec{r'}}|\rho_{\vec{l'},\vec{r'}}|^2\frac{1}{\rho_{\vec{l'}, \vec{r'}}}\frac{d\rho_{\vec{l'}, \vec{r'}}}{d\Theta_{\mu'}}\right],
	\end{aligned}
	\end{equation}
	where the right state $\vec{r}=(r_{0}, r_{1}, \cdots, r_{N})$ and the left state $\vec{l} = (l_{0}, l_{1}, \cdots, l_{N}) $ indicate a vector form of the $N$-particle state and we have $r(l)_{j} = \pm 1$ here. Then, it is easy to see that we can sampling the estimators  $(\frac{1}{\rho_{\vec{l}, \vec{r}}}\frac{d\rho_{\vec{l}, \vec{r}}}{d\Theta_\mu})^\ast\frac{(\mathcal{L}\rho)_{\vec{l},\vec{r}}}{\rho_{\vec{l},\vec{r}}}$, $(\frac{1}{\rho_{\vec{l}, \vec{r}}}\frac{d\rho_{\vec{l}, \vec{r}}}{d\Theta_\mu})^\ast$, $\frac{(\mathcal{L}\rho)_{\vec{l'},\vec{r'}}}{\rho_{\vec{l'},\vec{r'}}}$, $(\frac{1}{\rho_{\vec{l}, \vec{r}}}\frac{d\rho_{\vec{l}, \vec{r}}}{d\Theta_\mu})^\ast\frac{1}{\rho_{\vec{l}, \vec{r}}}\frac{d\rho_{\vec{l}, \vec{r}}}{d\Theta_\mu'}$, $(\frac{1}{\rho_{\vec{l}, \vec{r}}}\frac{d\rho_{\vec{l}, \vec{r}}}{d\Theta_\mu})^\ast$, and $\frac{1}{\rho_{\vec{l'}, \vec{r'}}}\frac{d\rho_{\vec{l'}, \vec{r'}}}{d\Theta_{\mu'}}$ through a Monte Carlo chain generated by the probability $|\rho_{\vec{l}, \vec{r}}|^2$. 
	
	To compute these estimators, we only need to consider three quantities $\rho_{\vec{l}, \vec{r}}$, $\frac{d\rho_{\vec{l},\vec{r}}}{d\Theta_\mu}$, and $(\mathcal{L}\rho)_{\vec{l},\vec{r}} = \langle\vec{l}|\mathcal{L}\rho|\vec{r}\rangle$. The first two quantities can be measured in our DQFNN. It is generally hard to obtain  $(\mathcal{L}\rho)_{\vec{l},\vec{r}}$, but here we emphasis that our Lindblad equation only contains the local operators $\sigma^z_j \sigma^z_{j+1}$,$\sigma^z_j \sigma^z_{j+2}$, $\sigma^x_j$, $\sigma^{+,-}_j$, and $\sigma^{+}_j \sigma^{-}_j$ , which satisfy the following relations:
	\begin{equation}
	\begin{aligned}
	&\left\langle\vec{l}\left|\sigma^{x}_{j} \right| \vec{r}\right\rangle =\delta_{r_{j},-l_{j}}  \prod_{k \neq j} \delta_{r_{k}, l_{k}}, \\
	&\left\langle\vec{l}\left|\sigma_{j}^{z} \sigma_{j+1}^{z}\right| \vec{r}\right\rangle =r_{j} r_{j+1} \delta_{\vec{l}, \vec{r}}, \\
	&\left\langle\vec{l}\left|\sigma_{j}^{z} \sigma_{j+2}^{z}\right| \vec{r}\right\rangle =r_{j} r_{j+2} \delta_{\vec{l}, \vec{r}},\\
	&\left\langle\vec{l}\left|\sigma_{j}^{+} \sigma_{j}^{-}\right| \vec{r}\right\rangle =\delta_{r_{j}, 1} \delta_{\vec{l}, \vec{r}}, \\
	&\left\langle\vec{l}\left|\sigma^{-}_{j}\right| \vec{r}\right\rangle =\delta_{r_{j}, 1} \delta_{l_{j},-1} \prod_{k \neq j} \delta_{l_{k}, r_{k}},\\
	&\left\langle\vec{l}\left|\sigma_{j}^{+}\right| \vec{r}\right\rangle =\delta_{l_{j},-1} \delta_{r_{j}, 1} \prod_{k \neq j} \delta_{l_{k}, r_{k}}.
	\end{aligned}
	\end{equation}
	Here we only consider the one dimensional case corresponding to  Eq.(9) in the main text. $(\mathcal{L}\rho)_{\vec{l},\vec{r}}$ can be written as:
	\begin{equation}
	\begin{aligned}
	\frac{(\mathcal{L}\rho)_{\vec{l},\vec{r}}}{\rho_{\vec{l},\vec{r}}}&=-\text{i}\frac{\langle\vec{l}|[H, \rho]|\vec{r}\rangle}{\rho_{\vec{l}, \vec{r}}}-\frac{\gamma}{2} \sum_{j}\left[\frac{\langle\vec{l}| \left\{\sigma_{j}^{+} \sigma_{j}^{-}, \rho\right\}|\vec{r}\rangle}{\rho_{\vec{l}, \vec{r}}}-2\frac{\langle\vec{l}|\sigma_{j}^{-} \rho \sigma_{j}^{+}|\vec{r}\rangle}{\rho_{\vec{l}, \vec{r}}}\right]\\
	&=\sum_{j=0}^{N}\lbrace -\text{i} J\left(l_{j} l_{j+1}-r_{j} r_{j+1}\right)-\text{i} h\left[\frac{\rho\left(\ldots,-l_{j} , \ldots\right)}{\rho\left(\ldots, l_{j},  \ldots\right)}-\frac{\rho\left(\ldots,-r_{j}, \ldots\right)}{\rho\left(\ldots, r_{j},  \ldots\right)}\right]\\
	&-\frac{\gamma}{2}\left(\delta_{l_{j}, 1}+\delta_{r_{j}, 1}\right)\rbrace-\gamma \delta_{l_{j},-1} \delta_{r_{j},-1} \frac{\rho\left(\ldots, l_{j-1}, 1, l_{j+1}, \ldots, \ldots, r_{j-1}, 1, r_{j+1}, \ldots\right)}{\rho\left(\ldots, l_{j-1},-1, l_{j+1}, \ldots, \ldots, r_{j-1},-1, r_{j+1}, \ldots\right)}.
	\end{aligned}
	\end{equation}
	
	To sample the $S$ and $\vec{f}$ efficiently, we generate a Markov chain of the right and left state configuration from step zero $(\vec{r}_0, \vec{l}_0) $ to step s $(\vec{r}_s, \vec{l}_s) $. In each step, a new trial right and left state configurations are randomly generated, and the new configuration is accepted with a probability $\min(1, |\frac{\rho_{\vec{l}_{\text{new}}, \vec{r}_{\text{new}}}}{\rho_{\vec{l}_{\text{old}}, \vec{r}_{\text{old}}}}|^2)$. We improve our sampling efficiency by generating each new configuration with slightly modifying the strategies discussed in \cite{CarleoOpenRBM, VinceOpenRBM}:
	
	1.  One lattice site is flipped both in the left state and right state. 
	
	2. One lattice site is flipped either in the left state or right state. 
	
	3. Two neighboring lattice sites are flipped either in the left state or right state. 
	
	4. All the lattice sites in the right state and left state are flipped.  
	
	5. A new random configuration of the right state and left state are generated.  
	
	In our numerical simulations, the probabilities of the first three types of moves are about $30$ times larger than the last two ones. To approximate the expectation value of a general hermitian observable, we consider the accepted probability as $\min(1, \frac{\rho_{\vec{l}_{\text{new}}, \vec{l}_{\text{new}}}}{\rho_{\vec{l}_{\text{old}}, \vec{l}_{\text{old}}}})$, and  generate a list of state configuration $\vec{l}_{0} \rightarrow \cdots \rightarrow \vec{l}_{s}$ by following possible moves:
	
	1.  One lattice site is flipped. 
	
	2. Two neighboring lattice sites are flipped. 
	
	3. All the lattice sites are flipped.  
	
	4. A new random configuration is generated.

	In our numerical simulations, the probabilities of the first two types of moves are about $30$ times larger than the last two types of moves.

	\section{Quantum updating scheme}
	
	With quantum devices, $S$ and $\vec{f}$ can be obtained efficiently from linear combinations of measurements of proper observables in the experiment.  A straightforward way is to do tomography of the output density matrix, which, however, is exponentially expensive as the system size increases. Thus, this method is limited to small systems. 
	To overcome this difficulty, we use the following methods to obtain $S$ and $\vec{f}$ efficiently. 
	We rewrite $S$ and $\vec{f}$ as follows:
	\begin{eqnarray}
	f_\mu &=& \text{Re}(\sum_{\vec{l}}\langle \vec{l}|\frac{\partial \rho}{\partial \Theta_\mu}\mathcal{L} \rho|\vec{l}\rangle - \sum_{\vec{l}}\langle \vec{l}|\rho \frac{\partial \rho}{\partial \Theta_\mu}|\vec{l}\rangle\sum_{\vec{l}'}\langle \vec{l}'|\rho \mathcal{L}\rho|\vec{l}'\rangle),\label{Eqf1}\\
	S_{\mu,\nu}&=&\text{Re}(\sum_{\vec{l}} \langle \vec{l}|\frac{\partial \rho}{\partial \Theta_\mu}\frac{\partial \rho}{\partial \Theta_\nu}|\vec{l}\rangle-\sum_{\vec{l}}\langle \vec{l} |\frac{\partial \rho}{\partial \Theta_\mu} \rho | \vec{l}\rangle\sum_{\vec{l}'}\langle \vec{l}' |\frac{\partial \rho}{\partial \Theta_\nu} \rho|\vec{l}' \rangle).\label{EqS1}
	\end{eqnarray}
	Thanks to the special structures of DQFNNs studied in this paper, the derivative $\frac{\partial \rho}{\partial \Theta_{\mu}}$ of $\rho$ with respect to the parameter $\Theta_{\mu}$  can be written as Eq.(10) in the main text. For the convenience, we write the $\sum_{\vec{l}}\langle \vec{l}| \cdots| \vec{l}\rangle$ as a trace operation and the Eqs. (\ref{Eqf1}-\ref{EqS1}) then reduce to:
	
	\begin{eqnarray}
	f_\mu &=&\frac{1}{2}\text{Re}\left\{ \text{Tr}(\rho(\Theta_\mu+\frac{\pi}{2})\mathcal{L}\rho) - \text{Tr}(\rho(\Theta_\mu-\frac{\pi}{2})\mathcal{L}\rho)-\left[\text{Tr}(\rho(\Theta_\mu+\frac{\pi}{2})\rho)- \text{Tr}(\rho(\Theta_\mu-\frac{\pi}{2})\rho) \right][\text{Tr}(\rho\mathcal{L}\rho)]  \right\} ,\label{Eqf2}\\
	S_{\mu, \nu} &=& \frac{1}{2}\text{Re} \left\{ \text{Tr}(\rho(\Theta_\mu+\frac{\pi}{2})\rho(\Theta_\nu+\frac{\pi}{2}))+
	\text{Tr}(\rho(\Theta_\mu-\frac{\pi}{2})\rho(\Theta_\nu+\frac{\pi}{2}))\right\}  \label{Eqs2} \\
	&+&\frac{1}{2}\text{Re} \left\{
	\text{Tr}(\rho(\Theta_\mu+\frac{\pi}{2})\rho(\Theta_\nu-\frac{\pi}{2}))
	+\text{Tr}(\rho(\Theta_\mu-\frac{\pi}{2})\rho(\Theta_\nu-\frac{\pi}{2})) \right\} \nonumber \\
	&-&\frac{1}{2}\text{Re} \left\{\left[\text{Tr}(\rho(\Theta_\mu+\frac{\pi}{2})\rho)- \text{Tr}(\rho(\Theta_\mu-\frac{\pi}{2})\rho) \right]\left[\text{Tr}(\rho(\Theta_\nu+\frac{\pi}{2})\rho)- \text{Tr}(\rho(\Theta_\nu-\frac{\pi}{2})\rho) \right]  \right\}.  \nonumber
	\end{eqnarray}

	As we can see in Eqs.(\ref{Eqf2}-\ref{Eqs2}), each term in $f_\mu$ and $S_{\mu,\nu}$ only has two general form, namely, $\text{Tr}[\rho(\Theta)\rho(\Theta')]$ and $\text{Tr}[\rho(\Theta)\mathcal{L}\rho(\Theta')]$, where $\Theta$ and $\Theta'$ are two parameter sets that can be different or identical.

	\begin{figure}
		\centering
		\includegraphics[scale=0.45]{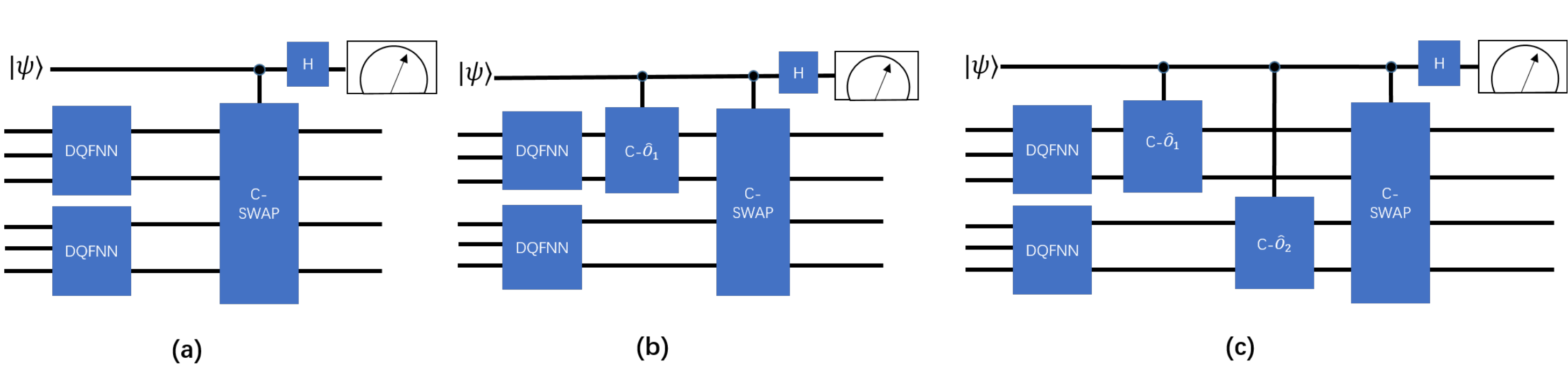}
		\caption{The quantum circuits for obtaining $S$ and $\vec{f}$ used in the DQFNN approach. Here, H denotes the Hardmard gate and C-SWAP represents the controlled-swap gate defined in Eq. (\ref{CSwapDef}). Each circuit is composited by two DQFNNs to generate $\rho(\Theta)$ and $\rho(\Theta')$ and one ancillary qubit for measuring.  (a) The circuit for measuring 	$\text{Tr}[\rho(\Theta)\rho(\Theta')]$, where $|\psi\rangle = (|0\rangle+|1\rangle)/\sqrt{2}$ (b) The circuit for measuring  $\text{Tr}[\rho(\Theta)\hat{O}_1\rho(\Theta')]$. C-$\hat{O}_1$ is the controlled-$\hat{O}_1$ gate defined in Eq.(\ref{controllO}), where $|\psi\rangle= (|0\rangle+|1\rangle)/\sqrt{2}$ for measuring the real part of the quantity and $|\psi\rangle = (|0\rangle+\text{i}|1\rangle)/\sqrt{2}$ for measuring the imaginary part of the quantity. (c) The circuit for measuring $\text{Tr}[\rho(\Theta)\hat{O}_1\rho(\Theta')\hat{O}_2]$, where $|\psi\rangle$ is initialized by the same strategy in (b). 
		}\label{fig:measurementcircuits}
	\end{figure}

	We adopt the quantum circuit in  Fig.\ref{fig:measurementcircuits}(a) to measure $\text{Tr}[\rho(\Theta)\rho(\Theta')]$, which has been discussed in \cite{Ekert2002lineardensitymatrix}. We firstly generate the output density matrices $\rho(\Theta)$ and $\rho(\Theta')$ from two DQFNNs, and introduce another ancillary qubit which is initialized to $(|0\rangle+|1\rangle)/\sqrt{2}$ state. We decompose the spectrum of the whole system: $\rho_{w} = \frac{1}{2}(|0\rangle\langle0|+|0\rangle\langle1|+|1\rangle\langle0|+|1\rangle\langle1|)\otimes\sum_{\vec{l},\vec{r},\vec{l}',\vec{r}'}\rho_{\vec{l},\vec{r}}\rho'_{\vec{l}',\vec{r}'}|\vec{l},\vec{l}'\rangle\langle \vec{r},\vec{r}'|$.
	
	After that, we apply a controlled-swap gate, which is defined as
	\begin{equation}
	U_{\text{C-SWAP}}|x\rangle|\psi\rangle_A|\phi\rangle_B = \left\{
	\begin{aligned} \label{CSwapDef}
	&|x\rangle|\psi\rangle_A|\phi\rangle_B, \;\text{ for } x = 0\\
	&|x\rangle|\phi\rangle_A|\psi\rangle_B, \;\text{ for } x = 1,
	\end{aligned}
	\right.
	\end{equation}
	on the whole system
	\begin{equation}
	\begin{aligned}
	U_{\text{C-SWAP}}\rho_{w}U_{\text{C-SWAP}}^\dagger = &\frac{1}{2}|0\rangle\langle0|\otimes\sum_{\vec{l},\vec{r},\vec{l}',\vec{r}'}\rho_{\vec{l},\vec{r}}\rho'_{\vec{l}',\vec{r}'}|\vec{l},\vec{l}'\rangle\langle \vec{r},\vec{r}'|\\
	&+\frac{1}{2}|0\rangle\langle1|\otimes\sum_{\vec{l},\vec{r},\vec{l}',\vec{r}'}\rho_{\vec{l},\vec{r}}\rho'_{\vec{l}',\vec{r}'}|\vec{l},\vec{l}'\rangle\langle \vec{r}',\vec{r}|\\
	&+\frac{1}{2}|1\rangle\langle0|\otimes\sum_{\vec{l},\vec{r},\vec{l}',\vec{r}'}\rho_{\vec{l},\vec{r}}\rho'_{\vec{l}',\vec{r}'}|\vec{l}',\vec{l}\rangle\langle \vec{r},\vec{r}'|\\
	&+\frac{1}{2}|1\rangle\langle1|\otimes\sum_{\vec{l},\vec{r},\vec{l}',\vec{r}'}\rho_{\vec{l},\vec{r}}\rho'_{\vec{l}',\vec{r}'}|\vec{l}',\vec{l}\rangle\langle \vec{r}',\vec{r}|.
	\end{aligned}
	\end{equation}
	Finally, we apply another Hardmard gate on the ancilla qubit and measure the spin probability in state $|0\rangle$:
	\begin{equation}
	\begin{aligned}
	P(0) &= \frac{1}{2}+\frac{1}{4}\sum_{\vec{l},\vec{r},\vec{l}',\vec{r}',\vec{m},\vec{n}}
	\rho_{\vec{l},\vec{r}}\rho'_{\vec{l}',\vec{r}'}\langle \vec{m},\vec{n}|\vec{l},\vec{l}'\rangle\langle \vec{r}',\vec{r}|\vec{m},\vec{n}\rangle+\frac{1}{4}\sum_{\vec{l},\vec{r},\vec{l}',\vec{r}',\vec{m},\vec{n}}
	\rho_{\vec{l},\vec{r}}\rho'_{\vec{l}',\vec{r}'}\langle \vec{m},\vec{n}|\vec{l}',\vec{l}\rangle\langle \vec{r},\vec{r}'|\vec{m},\vec{n}\rangle\\
	&=\frac{1}{2}+\frac{1}{4}\sum_{\vec{l},\vec{r},\vec{l}',\vec{r}',\vec{m},\vec{n}}\rho_{\vec{l},\vec{r}}\rho'_{\vec{l}',\vec{r}'}\delta_{\vec{m},\vec{l}}\delta_{\vec{n},\vec{l}'}\delta_{\vec{m},\vec{r}'}\delta_{\vec{n},\vec{r}}
	+\frac{1}{4}\sum_{\vec{l},\vec{r},\vec{l}',\vec{r}',\vec{m},\vec{n}}
	\rho_{\vec{l},\vec{r}}\rho'_{\vec{l}',\vec{r}'}\delta_{\vec{m},\vec{l}'}\delta_{\vec{n},\vec{l}}\delta_{\vec{m},\vec{r}}\delta_{\vec{n},\vec{r}'}\\
	& = \frac{1}{2}+\frac{1}{2}\text{Tr}\left[\rho(\Theta)\rho(\Theta')\right],
	\end{aligned}
	\end{equation}
	where $P(0)$ is the probability that the ancillary qubit is in state $|0\rangle$, so that we can obtain $\text{Tr}\left[\rho(\Theta)\rho(\Theta')\right]$ via measure the ancillary state population. 
	
	To measuring $\text{Tr}[\rho(\Theta)\mathcal{L}\rho(\Theta')]$, we expand the Lindblad equation into a explicit form:
	\begin{equation}
	\begin{aligned}
	\text{Tr}[\rho(\Theta)\mathcal{L}\rho(\Theta')] &= -\text{i} \text{Tr}\left[\rho(\Theta)H\rho(\Theta')\right]+\text{i}\text{Tr}\left[\rho(\Theta')H\rho(\Theta)\right]\\
	&-\frac{\gamma}{2}\sum_i\left\{\text{Tr}\left[\rho(\Theta)\sigma_{i}^{+}\sigma_{i}^{-}\rho(\Theta')\right]+\text{Tr}\left[\rho(\Theta')\sigma_{i}^{+}\sigma_{i}^{-}\rho(\Theta)\right]-2\text{Tr}\left[\sigma_{i}^-\rho(\Theta) \sigma_{i}^+\rho(\Theta')\right]\right\}.
	\end{aligned}
	\end{equation}
	Taking into account the one dimensional case in main text Eq.(9), the Hamiltonian $H$ only involves the local Pauli operators, so that we only need to consider $\text{Tr}\left[\rho(\Theta)\sigma_i^x\rho(\Theta')\right]$ and $\text{Tr}\left[\rho(\Theta)\sigma_{i}^z\sigma_{i+1}^{z}\rho(\Theta')\right]$. Here, we write the local Pauli operators in a uniform form $\hat{O}=\sigma_i^x, \sigma_{i}^z\sigma_{i+1}^{z} $. Utilizing the quantum circuits in Fig.\ref{fig:measurementcircuits}(b), we can measure the real and imaginary part of $\text{Tr}\left\{\rho(\Theta)\hat{O}\rho(\Theta')\right\}$ separately by setting the initial state of the ancillary qubit as $(|0\rangle+|1\rangle)/\sqrt{2}$ for measuring real part and $(|0\rangle+\text{i}|1\rangle)/\sqrt{2}$ for measuring the imaginary part. Then, we introduce a controlled-$\hat{O}$ gate, which is defined as:
	\begin{equation}
	\label{controllO}
	\hat{O}^c |x\rangle |\psi\rangle =  |x\rangle \hat{O}^x|\psi\rangle.
	\end{equation}
	We apply the controlled-$\hat{O}$ gate between the first subsystem and the ancillary qubit and apply another controlled-swap gate as: 
	\begin{equation}
	\begin{aligned}
	U_{\text{C-SWAP}}\hat{O}^c\rho_{w}^{\text{Re}}(\hat{O}^c)^\dagger U_{\text{C-SWAP}}^\dagger = &\frac{1}{2}|0\rangle\langle0|\otimes\sum_{\vec{l},\vec{r},\vec{l}',\vec{r}'}\rho_{\vec{l},\vec{r}}\rho'_{\vec{l}',\vec{r}'}|\vec{l},\vec{l}'\rangle\langle \vec{r},\vec{r}'|\\
	&+\frac{1}{2}|0\rangle\langle1|\otimes\sum_{\vec{l},\vec{r},\vec{l}',\vec{r}'}\rho_{\vec{l},\vec{r}}\rho'_{\vec{l}',\vec{r}'}|\vec{l},\vec{l}'\rangle\langle \vec{r}',\vec{r}|\hat{O}^\dagger\\
	&+\frac{1}{2}|1\rangle\langle0|\otimes\sum_{\vec{l},\vec{r},\vec{l}',\vec{r}'}\rho_{\vec{l},\vec{r}}\rho'_{\vec{l}',\vec{r}'}\hat{O}|\vec{l}',\vec{l}\rangle\langle \vec{r},\vec{r}'|\\
	&+\frac{1}{2}|1\rangle\langle1|\otimes\sum_{\vec{l},\vec{r},\vec{l}',\vec{r}'}\rho_{\vec{l},\vec{r}}\rho'_{\vec{l}',\vec{r}'}\hat{O}|\vec{l}',\vec{l}\rangle\langle \vec{r}',\vec{r}|\hat{O}^\dagger,
	\end{aligned}
	\end{equation}
	and 
	\begin{equation}
	\begin{aligned}
	U_{\text{C-SWAP}}\hat{O}^c\rho_{w}^{\text{Im}}(\hat{O}^c)^\dagger U_{\text{C-SWAP}}^\dagger = &\frac{1}{2}|0\rangle\langle0|\otimes\sum_{\vec{l},\vec{r},\vec{l}',\vec{r}'}\rho_{\vec{l},\vec{r}}\rho'_{\vec{l}',\vec{r}'}|\vec{l},\vec{l}'\rangle\langle \vec{r},\vec{r}'|\\
	&-\frac{\text{i}}{2}|0\rangle\langle1|\otimes\sum_{\vec{l},\vec{r},\vec{l}',\vec{r}'}\rho_{\vec{l},\vec{r}}\rho'_{\vec{l}',\vec{r}'}|\vec{l},\vec{l}'\rangle\langle \vec{r}',\vec{r}|\hat{O}^\dagger\\
	&+\frac{\text{i}}{2}|1\rangle\langle0|\otimes\sum_{\vec{l},\vec{r},\vec{l}',\vec{r}'}\rho_{\vec{l},\vec{r}}\rho'_{\vec{l}',\vec{r}'}\hat{O}|\vec{l}',\vec{l}\rangle\langle \vec{r},\vec{r}'|\\
	&+\frac{1}{2}|1\rangle\langle1|\otimes\sum_{\vec{l},\vec{r},\vec{l}',\vec{r}'}\rho_{\vec{l},\vec{r}}\rho'_{\vec{l}',\vec{r}'}\hat{O}|\vec{l}',\vec{l}\rangle\langle \vec{r}',\vec{r}|\hat{O}^\dagger.
	\end{aligned}
	\end{equation}
	
	After apply a Hardmard gate on the ancillary qubit, we measure the spin population of ancillary qubit
	\begin{equation}
	\begin{aligned}
	P(0)_{\text{Re}} &= \frac{1}{4}+
	\frac{1}{4}\sum_{\vec{l},\vec{r},\vec{l}',\vec{r}',\vec{m},\vec{n}}
	\rho_{\vec{l},\vec{r}}\rho'_{\vec{l}',\vec{r}'}\langle\vec{m},\vec{n}|\vec{l},\vec{l}'\rangle\langle \vec{r}|\vec{n}\rangle \langle \vec{r}'|\hat{O}^\dagger|\vec{m}\rangle
	+\frac{1}{4}\sum_{\vec{l},\vec{r},\vec{l}',\vec{r}',\vec{m},\vec{n}}
	\rho_{\vec{l},\vec{r}}\rho'_{\vec{l}',\vec{r}'}\langle \vec{m}|\hat{O}|\vec{l}'\rangle
	\langle\vec{n}|\vec{l}\rangle
	\langle \vec{r}, \vec{r}'|\vec{m},\vec{n}\rangle\\
	&+\frac{1}{4}\sum_{\vec{l},\vec{r},\vec{l}',\vec{r}',\vec{m},\vec{n}}
	\rho_{\vec{l},\vec{r}}\rho'_{\vec{l}',\vec{r}'}\langle \vec{m}|\hat{O}|\vec{l}'\rangle\langle\vec{n}|\vec{l}\rangle\langle \vec{r}|\vec{n}\rangle
	\langle\vec{r}'|\hat{O}^\dagger|\vec{m}\rangle
	\\
	&=\frac{1}{4}+\frac{1}{4}\sum_{\vec{l},\vec{r},\vec{l}',\vec{r}',\vec{m},\vec{n}}\rho_{\vec{l},\vec{r}}\rho'_{\vec{l}',\vec{r}'}\hat{O}^\dagger_{\vec{r}',\vec{m}}\delta_{\vec{n},\vec{l}'}\delta_{\vec{m},\vec{l}}\delta_{\vec{n},\vec{r}}
	+\frac{1}{4}\sum_{\vec{l},\vec{r},\vec{l}',\vec{r}',\vec{m},\vec{n}}
	\rho_{\vec{l},\vec{r}}\rho'_{\vec{l}',\vec{r}'}\hat{O}_{\vec{m},\vec{l}'}\delta_{\vec{n},\vec{l}}\delta_{\vec{m},\vec{r}}\delta_{\vec{n},\vec{r}'}\\
	&+\frac{1}{4}\sum_{\vec{l},\vec{r},\vec{l}',\vec{r}',\vec{m},\vec{n}}
	\rho_{\vec{l},\vec{r}}\rho'_{\vec{l}',\vec{r}'}\hat{O}_{\vec{m},\vec{l}'}\hat{O}^\dagger_{\vec{r}',\vec{m}}\delta_{\vec{n},\vec{l}}\delta_{\vec{n},\vec{r}}
	\\
	& = \frac{1}{4}+\frac{1}{4}\text{Tr}\left[\rho(\Theta)\hat{O}\rho(\Theta')\right]+\frac{1}{4}\text{Tr}\left[\rho(\Theta')\hat{O}^\dagger\rho(\Theta)\right]+\frac{1}{4}\text{Tr}\left[\rho(\Theta)\right]\left[\hat{O}\rho(\Theta')\hat{O}^\dagger\right]\\
	& = \frac{1}{2}+\frac{1}{2}\text{Re}\left\{\text{Tr}\left[\rho(\Theta)\hat{O}\rho(\Theta')\right]\right\},
	\end{aligned}
	\end{equation}
	and 
	\begin{equation}
	\begin{aligned}
	P(0)_{\text{Im}} &= \frac{1}{4}-\frac{\text{i}}{4}\sum_{\vec{l},\vec{r},\vec{l}',\vec{r}',\vec{m},\vec{n}}
	\rho_{\vec{l},\vec{r}}\rho'_{\vec{l}',\vec{r}'}\langle\vec{m},\vec{n}|\vec{l},\vec{l}'\rangle\langle \vec{r}|\vec{n}\rangle \langle \vec{r}'|\hat{O}^\dagger|\vec{m}\rangle
	+\frac{\text{i}}{4}\sum_{\vec{l},\vec{r},\vec{l}',\vec{r}',\vec{m},\vec{n}}
	\rho_{\vec{l},\vec{r}}\rho'_{\vec{l}',\vec{r}'}\langle \vec{m}|\hat{O}|\vec{l}'\rangle
	\langle\vec{n}|\vec{l}\rangle
	\langle \vec{r}, \vec{r}'|\vec{m},\vec{n}\rangle\\
	&+\frac{1}{4}\sum_{\vec{l},\vec{r},\vec{l}',\vec{r}',\vec{m},\vec{n}}
	\rho_{\vec{l},\vec{r}}\rho'_{\vec{l}',\vec{r}'}\langle \vec{m}|\hat{O}|\vec{l}'\rangle\langle\vec{n}|\vec{l}\rangle\langle \vec{r}|\vec{n}\rangle
	\langle\vec{r}'|\hat{O}^\dagger|\vec{m}\rangle
	\\
	&=\frac{1}{4}-\frac{\text{i}}{4}\sum_{\vec{l},\vec{r},\vec{l}',\vec{r}',\vec{m},\vec{n}}\rho_{\vec{l},\vec{r}}\rho'_{\vec{l}',\vec{r}'}\hat{O}^\dagger_{\vec{r}',\vec{m}}\delta_{\vec{n},\vec{l}'}\delta_{\vec{m},\vec{l}}\delta_{\vec{n},\vec{r}}
	+\frac{\text{i}}{4}\sum_{\vec{l},\vec{r},\vec{l}',\vec{r}',\vec{m},\vec{n}}
	\rho_{\vec{l},\vec{r}}\rho'_{\vec{l}',\vec{r}'}\hat{O}_{\vec{m},\vec{l}'}\delta_{\vec{n},\vec{l}}\delta_{\vec{m},\vec{r}}\delta_{\vec{n},\vec{r}'}\\
	&+\frac{1}{4}\sum_{\vec{l},\vec{r},\vec{l}',\vec{r}',\vec{m},\vec{n}}
	\rho_{\vec{l},\vec{r}}\rho'_{\vec{l}',\vec{r}'}\hat{O}_{\vec{m},\vec{l}'}\hat{O}^\dagger_{\vec{r}',\vec{m}}\delta_{\vec{n},\vec{l}}\delta_{\vec{n},\vec{r}}
	\\
	& = \frac{1}{4}+\frac{\text{i}}{4}\text{Tr}\left[\rho(\Theta)\hat{O}\rho(\Theta')\right]-\frac{\text{i}}{4}\text{Tr}\left[\rho(\Theta')\hat{O}^\dagger\rho(\Theta)\right]+\frac{1}{4}\text{Tr}\left[\rho(\Theta)\right]\left[\hat{O}\rho(\Theta')\hat{O}^\dagger\right]\\
	& = \frac{1}{2}-\frac{1}{2}\text{Im}\left\{\text{Tr}\left[\rho(\Theta)\hat{O}\rho(\Theta')\right]\right\},
	\end{aligned}
	\end{equation}
	and obtain $\text{Tr}\left[\rho(\Theta)\hat{O}\rho(\Theta')\right] = 2P(0)_{\text{Re}}-1+\text{i}(1-2P(0)_{\text{Im}})$.

	To calculate the term $\text{Tr}\left[\rho(\Theta)\sigma_{i}^{+}\sigma_{i}^{-}\rho(\Theta')\right]$, we divide  it into $\text{Tr}\left[\rho(\Theta)\rho(\Theta')\right]$ and $\text{Tr}\left[\rho(\Theta)\sigma_i^z\rho(\Theta')\right]$ by using $\sigma_i^+\sigma_i^- = \frac{1}{2}+\frac{1}{2}\sigma_i^z$. These two quantites can be measured by using the method above. We expand the last term $\sigma_i^-\rho\sigma_i^+ = \sigma_i^x\rho\sigma_i^x+\sigma_i^y\rho\sigma_i^y-\text{i}(\sigma_i^y\rho\sigma_i^x-\sigma_i^x\rho\sigma_i^y)$ and measure the real part and imaginary part separately. To calculate $\text{Tr}\left[\rho(\Theta)\sigma_i^{x,y}\rho(\Theta')\sigma_i^{x,y}\right]$, as shown in Fig.\ref{fig:measurementcircuits}(c), we apply two controlled-$\hat{O}_{1,2}$ gates, where $\hat{O}_{1,2} =\sigma_i^{x,y}$, and apply the controlled-swap gate
	\begin{equation}
	\begin{aligned}
	U_{\text{C-SWAP}}\hat{O}_2^c\hat{O}_1^c\rho_{w}^{Re}(\hat{O}_1^c)^\dagger(\hat{O}_2^c)^\dagger U_{\text{C-SWAP}}^\dagger = &\frac{1}{2}|0\rangle\langle0|\otimes\sum_{\vec{l},\vec{r},\vec{l}',\vec{r}'}\rho_{\vec{l},\vec{r}}\rho'_{\vec{l}',\vec{r}'}|\vec{l},\vec{l}'\rangle\langle \vec{r},\vec{r}'|\\
	&+\frac{1}{2}|0\rangle\langle1|\otimes\sum_{\vec{l},\vec{r},\vec{l}',\vec{r}'}\rho_{\vec{l},\vec{r}}\rho'_{\vec{l}',\vec{r}'}|\vec{l},\vec{l}'\rangle\langle \vec{r}',\vec{r}|\hat{O}_1^\dagger\hat{O}_2^\dagger\\
	&+\frac{1}{2}|1\rangle\langle0|\otimes\sum_{\vec{l},\vec{r},\vec{l}',\vec{r}'}\rho_{\vec{l},\vec{r}}\rho'_{\vec{l}',\vec{r}'}\hat{O}_1 \hat{O}_2|\vec{l}',\vec{l}\rangle\langle \vec{r},\vec{r}'|\\
	&+\frac{1}{2}|1\rangle\langle1|\otimes\sum_{\vec{l},\vec{r},\vec{l}',\vec{r}'}\rho_{\vec{l},\vec{r}}\rho'_{\vec{l}',\vec{r}'}\hat{O}_1 \hat{O}_2|\vec{l}',\vec{l}\rangle\langle \vec{r}',\vec{r}|\hat{O}_1^\dagger\hat{O}_2^\dagger,
	\end{aligned}
	\end{equation}
	and 
	\begin{equation}
	\begin{aligned}
	U_{\text{C-SWAP}}\hat{O}_2^c\hat{O}_1^c\rho_{w}^{Im}(\hat{O}_1^c)^\dagger(\hat{O}_2^c)^\dagger U_{\text{C-SWAP}}^\dagger = &\frac{1}{2}|0\rangle\langle0|\otimes\sum_{\vec{l},\vec{r},\vec{l}',\vec{r}'}\rho_{\vec{l},\vec{r}}\rho'_{\vec{l}',\vec{r}'}|\vec{l},\vec{l}'\rangle\langle \vec{r},\vec{r}'|\\
	&-\frac{\text{i}}{2}|0\rangle\langle1|\otimes\sum_{\vec{l},\vec{r},\vec{l}',\vec{r}'}\rho_{\vec{l},\vec{r}}\rho'_{\vec{l}',\vec{r}'}|\vec{l},\vec{l}'\rangle\langle \vec{r}',\vec{r}|\hat{O}_1^\dagger\hat{O}_2^\dagger\\
	&+\frac{\text{i}}{2}|1\rangle\langle0|\otimes\sum_{\vec{l},\vec{r},\vec{l}',\vec{r}'}\rho_{\vec{l},\vec{r}}\rho'_{\vec{l}',\vec{r}'}\hat{O}_1 \hat{O}_2|\vec{l}',\vec{l}\rangle\langle \vec{r},\vec{r}'|\\
	&+\frac{1}{2}|1\rangle\langle1|\otimes\sum_{\vec{l},\vec{r},\vec{l}',\vec{r}'}\rho_{\vec{l},\vec{r}}\rho'_{\vec{l}',\vec{r}'}\hat{O}_1 \hat{O}_2|\vec{l}',\vec{l}\rangle\langle \vec{r}',\vec{r}|\hat{O}_1^\dagger\hat{O}_2^\dagger.
	\end{aligned}
	\end{equation}
	Then, similar to our previous discussion, we measure the ancillary qubit 
	
	\begin{equation}
	\begin{aligned}
	P(0)_{\text{Re}} &= \frac{1}{4}+
	\frac{1}{4}\sum_{\vec{l},\vec{r},\vec{l}',\vec{r}',\vec{m},\vec{n}}
	\rho_{\vec{l},\vec{r}}\rho'_{\vec{l}',\vec{r}'}\langle\vec{m},\vec{n}|\vec{l},\vec{l}'\rangle\langle \vec{r}|\hat{O}_2^\dagger|\vec{n}\rangle \langle \vec{r}'|\hat{O}_1^\dagger|\vec{m}\rangle
	+\frac{1}{4}\sum_{\vec{l},\vec{r},\vec{l}',\vec{r}',\vec{m},\vec{n}}
	\rho_{\vec{l},\vec{r}}\rho'_{\vec{l}',\vec{r}'}\langle \vec{m}|\hat{O}_1|\vec{l}'\rangle
	\langle\vec{n}|\hat{O}_2|\vec{l}\rangle
	\langle \vec{r}, \vec{r}'|\vec{m},\vec{n}\rangle\\
	&+\frac{1}{4}\sum_{\vec{l},\vec{r},\vec{l}',\vec{r}',\vec{m},\vec{n}}
	\rho_{\vec{l},\vec{r}}\rho'_{\vec{l}',\vec{r}'}\langle \vec{m}|\hat{O}_1|\vec{l}'\rangle\langle\vec{n}|\hat{O}_2|\vec{l}\rangle\langle \vec{r}|\hat{O}_2^\dagger|\vec{n}\rangle
	\langle\vec{r}'|\hat{O}_1^\dagger|\vec{m}\rangle
	\\
	&=\frac{1}{4}+\frac{1}{4}\sum_{\vec{l},\vec{r},\vec{l}',\vec{r}',\vec{m},\vec{n}}\rho_{\vec{l},\vec{r}}\rho'_{\vec{l}',\vec{r}'}(\hat{O}_1^\dagger)_{\vec{r}',\vec{m}}(\hat{O}_2^\dagger)_{\vec{r},\vec{n}}\delta_{\vec{n},\vec{l}'}\delta_{\vec{m},\vec{l}}
	+\frac{1}{4}\sum_{\vec{l},\vec{r},\vec{l}',\vec{r}',\vec{m},\vec{n}}
	\rho_{\vec{l},\vec{r}}\rho'_{\vec{l}',\vec{r}'}(\hat{O}_1)_{\vec{m},\vec{l}'}(\hat{O}_2)_{\vec{n},\vec{l}}\delta_{\vec{m},\vec{r}}\delta_{\vec{n},\vec{r}'}\\
	&+\frac{1}{4}\sum_{\vec{l},\vec{r},\vec{l}',\vec{r}',\vec{m},\vec{n}}
	\rho_{\vec{l},\vec{r}}\rho'_{\vec{l}',\vec{r}'}(\hat{O}_1)_{\vec{m},\vec{l}'}(\hat{O}^\dagger_1)_{\vec{r}',\vec{m}}(\hat{O}_2)_{\vec{n},\vec{l}}(\hat{O}_2^\dagger)_{\vec{r},\vec{n}}
	\\
	& = \frac{1}{4}+\frac{1}{4}\text{Tr}\left[\rho(\Theta)\hat{O}_1\rho(\Theta')\hat{O}_2\right]+\frac{1}{4}\text{Tr}\left[\rho(\Theta')\hat{O}_1^\dagger\rho(\Theta)\hat{O}_2^\dagger\right]+\frac{1}{4}\text{Tr}\left[\hat{O}_2\rho(\Theta)\hat{O}_2^\dagger\right]\left[\hat{O}_1\rho(\Theta')\hat{O}_1^\dagger\right]\\
	& = \frac{1}{2}+\frac{1}{2}\text{Re}\left\{\text{Tr}\left[\rho(\Theta)\hat{O}\rho(\Theta')\right]\right\},
	\end{aligned}
	\end{equation}
	and 
	\begin{equation}
	\begin{aligned}
	P(0)_{\text{Im}} &= \frac{1}{4}-
	\frac{\text{i}}{4}\sum_{\vec{l},\vec{r},\vec{l}',\vec{r}',\vec{m},\vec{n}}
	\rho_{\vec{l},\vec{r}}\rho'_{\vec{l}',\vec{r}'}\langle\vec{m},\vec{n}|\vec{l},\vec{l}'\rangle\langle \vec{r}|\hat{O}_2^\dagger|\vec{n}\rangle \langle \vec{r}'|\hat{O}_1^\dagger|\vec{m}\rangle
	+\frac{\text{i}}{4}\sum_{\vec{l},\vec{r},\vec{l}',\vec{r}',\vec{m},\vec{n}}
	\rho_{\vec{l},\vec{r}}\rho'_{\vec{l}',\vec{r}'}\langle \vec{m}|\hat{O}_1|\vec{l}'\rangle
	\langle\vec{n}|\hat{O}_2|\vec{l}\rangle
	\langle \vec{r}, \vec{r}'|\vec{m},\vec{n}\rangle\\
	&+\frac{1}{4}\sum_{\vec{l},\vec{r},\vec{l}',\vec{r}',\vec{m},\vec{n}}
	\rho_{\vec{l},\vec{r}}\rho'_{\vec{l}',\vec{r}'}\langle \vec{m}|\hat{O}_1|\vec{l}'\rangle\langle\vec{n}|\hat{O}_2|\vec{l}\rangle\langle \vec{r}|\hat{O}_2^\dagger|\vec{n}\rangle
	\langle\vec{r}'|\hat{O}_1^\dagger|\vec{m}\rangle
	\\
	&=\frac{1}{4}-\frac{\text{i}}{4}\sum_{\vec{l},\vec{r},\vec{l}',\vec{r}',\vec{m},\vec{n}}\rho_{\vec{l},\vec{r}}\rho'_{\vec{l}',\vec{r}'}(\hat{O}_1^\dagger)_{\vec{r}',\vec{m}}(\hat{O}_2^\dagger)_{\vec{r},\vec{n}}\delta_{\vec{n},\vec{l}'}\delta_{\vec{m},\vec{l}}
	+\frac{\text{i}}{4}\sum_{\vec{l},\vec{r},\vec{l}',\vec{r}',\vec{m},\vec{n}}
	\rho_{\vec{l},\vec{r}}\rho'_{\vec{l}',\vec{r}'}(\hat{O}_1)_{\vec{m},\vec{l}'}(\hat{O}_2)_{\vec{n},\vec{l}}\delta_{\vec{m},\vec{r}}\delta_{\vec{n},\vec{r}'}\\
	&+\frac{1}{4}\sum_{\vec{l},\vec{r},\vec{l}',\vec{r}',\vec{m},\vec{n}}
	\rho_{\vec{l},\vec{r}}\rho'_{\vec{l}',\vec{r}'}(\hat{O}_1)_{\vec{m},\vec{l}'}(\hat{O}^\dagger_1)_{\vec{r}',\vec{m}}(\hat{O}_2)_{\vec{n},\vec{l}}(\hat{O}_2^\dagger)_{\vec{r},\vec{n}}
	\\
	& = \frac{1}{4}+\frac{\text{i}}{4}\text{Tr}\left[\rho(\Theta)\hat{O}_1\rho(\Theta')\hat{O}_2\right]-\frac{\text{i}}{4}\text{Tr}\left[\rho(\Theta')\hat{O}_1^\dagger\rho(\Theta)\hat{O}_2^\dagger\right]+\frac{1}{4}\text{Tr}\left[\hat{O}_2\rho(\Theta)\hat{O}_2^\dagger\right]\left[\hat{O}_1\rho(\Theta')\hat{O}_1^\dagger\right]\\
	& = \frac{1}{2}-\frac{1}{2}\text{Im}\left\{\text{Tr}\left[\rho(\Theta)\hat{O}\rho(\Theta')\right]\right\}.
	\end{aligned}
	\end{equation}
	
	and obtain $\text{Tr}\left[\rho(\Theta)\hat{O}_1\rho(\Theta')\hat{O}_2\right] = 2P(0)_{\text{Re}}-1+\text{i}(1-2P(0)_{\text{Im}})$.
	
	It is easy to see that $\text{Tr}[\rho(\Theta)\hat{O}\rho(\Theta)]$ and $\text{Tr}[\rho(\Theta)\hat{O}_1\rho(\Theta)\hat{O}_2]$ is a real number, so that it can be measured by the quantum circuits in Fig.\ref{fig:measurementcircuits}(b) and (c) respectively by setting the initial ancillary qubit state as $(|0\rangle+|1\rangle)/\sqrt{2}$. In summary, by using the techniques discussed above we can obtain $S$ and $\vec{f}$ efficiently with quantum devices.

	\section{The Training Scheme}
	
	We use the following procedures to train the DQFNN for solving  the dynamics and stationary states for open quantum systems:	
	
	\uppercase\expandafter{\romannumeral1} \quad DQFNN initialization: In our case, 
	only the single qubit gate is parameterized by $R(\vec{\theta}) = e^{i \theta_1 \sigma_z/2}e^{i \theta_2 \sigma_x/2}e^{i \theta_3 \sigma_z/2} $. At the beginning, we initialize the DQFNN by assigning small random numbers to all the parameters, so that each single qubit gate is approach an identity matrix.
	
	\uppercase\expandafter{\romannumeral2} \quad Feedforward:
	
	\uppercase\expandafter{\romannumeral2}.1. Set $i=1$, and set the state of qubits in the input layer to $\rho^i_o = (\otimes_{k=1}^{N_{1}}|+\rangle_{k})(\otimes_{k=1}^{N_{1}}\langle+|_{k})$, where $N_1$ is the number of qubits in input layer, and $|+\rangle_{k} = (|0\rangle+|1\rangle)/\sqrt{2}$.
	
	\uppercase\expandafter{\romannumeral2}.2. Set the state of qubits in the $(i+1)$-th layer to $\rho^{i+1}_{\text{in}} = (\otimes_{k=1}^{N_{i+1}}|+\rangle_{k})(\otimes_{k=1}^{N_{i+1}}\langle+|_{k})$, where $N_{i+1}$ is the number of qubits in $(i+1)$-th layer. 
	
	\uppercase\expandafter{\romannumeral2}.3. Apply unitary operations between $i$-th layer and $(i+1)$-th layer. 
	\begin{equation}
	\rho^{i+1} = 
	U_{i}^{w_i}  \ldots U_{i}^{j}  \ldots U_{i}^{1}\left(\rho^{i}_o\otimes \rho^{i+1}_{\text{in}}\right) {U_{i}^{1}}^{\dagger} \ldots {U_{i}^{j}}^{\dagger} \ldots {U_{i}^{w_i}}^{\dagger},
	\end{equation}  
	where $w_i$ is the number of perceptrons between $i$-th layer and ($(i+1)$)-th layer.

	\uppercase\expandafter{\romannumeral2}.4. Partial trace the $i$-th layer, namely, $\rho^{i+1}_o=\text{Tr}_{i}(\rho^{i+1})$ and set $i$ = $i+1$.
	
	\uppercase\expandafter{\romannumeral2}.5. Reuse the qubits. If $i<d$, the partial traced qubits in $i$-th layer is re-initialized as $(|0\rangle+|1\rangle)/\sqrt{2}$ and they can be used in the rest of the network.
	
	\uppercase\expandafter{\romannumeral2}.6. Repeat \uppercase\expandafter{\romannumeral2}.2-\uppercase\expandafter{\romannumeral2}.4 until $i=d$.

	\uppercase\expandafter{\romannumeral3} \quad Update parameters:
	
	\uppercase\expandafter{\romannumeral3}.1. Compute $S$ and $\vec{f}$. As the discussion in section \uppercase\expandafter{\romannumeral1} and section \uppercase\expandafter{\romannumeral2}, we can obtain the $S$ and $\vec{f}$ via a stochastic Markov-chain sampling or directly measure these two quantities. 
	
	\uppercase\expandafter{\romannumeral3}.2. Update parameters. According to the system equation in the main text, the parameters can be updated as $\Theta_\mu^{\text{new}} = \Theta_\mu^{\text{old}} + \lambda \sum_{\nu} (S^{-1})_{\mu,\nu} f_{\nu}$. Here we choose  $\lambda$ to be small enough to monitor the convergence of the optimization process. 
	
	\uppercase\expandafter{\romannumeral4} \quad Repeat step \uppercase\expandafter{\romannumeral2} and \uppercase\expandafter{\romannumeral3} until $\delta \mathcal{L}$ reaches minimal and the process get converged.

	\section{Total Number of Parameters and Symmetry Implementation Scheme}
	
	In our DQFNN, the $j$-th perceptron in $i$-th layer contains three qubits and a series of coupling unitary operation belongs to $SU(2^3)$ group with the most universal form is:
	\begin{equation}
	U_{i,j} = \exp\left[-\text{i}\sum_{\alpha_1, \alpha_2, \alpha_3} \frac{\Theta_{i,j,\vec{\alpha}}}{2} \sigma^{\alpha_1}\otimes\sigma^{\alpha_2}\otimes\sigma^{\alpha_3}\right],
	\end{equation}
	where $\Theta_{i,j,\vec{\alpha}}$ denote the parameters for the $j$-th perceptron in the $i$-th layer.  The index $\alpha_{1,2,3}$ takes $0, x, y, z$ to indicate the pauli matrix $I, \sigma^x, \sigma^y, \sigma^z$ respectly. Hence, to fully describe a quantum perceptron in this case needs $64$ parameters at most. As mentioned in the main text, we use an experimentally more practical quantum circuit antasz, which contains $12$ Euler rotations and $6$ CNOT gates, to implement a quantum perceptron with only $36$ parameters. 
	%

	For a fully connected DQFNN, the perceptrons are composed by any possible combination of two qubits from the $i$-th layer and one qubit from the $(i+1)$-th layer, where the total number of perceptrons between two neighbor layers is $C_{n_{i}}^{2}n_{i+1}$. Hence, the total number of parameters of the DQFNN is:
	\begin{equation}
	\sum_{i=1}^{d-1}C_{n_{i}}^{2}n_{i+1}\times N_p \approx O(N_{\text{All}}^3)
	\end{equation}
	where $d$ is the total number of layers, $n_i$ denotes the number of qubits in the $i$-th layer, $N_p$ is the number of parameters for each perceptron, and $N_{\text{All}}$ denotes the total number of qubits in the DQFNN.  
	Here we reduce the number of parameters by using the symmetry structure in the Hamiltonian we discussed in our main text. We restrict each perceptron to only involve local connection between $i$-th and $(i+1)$-th layers.  The $j$-th qubit in the $(i+1)$-th layer is only grouped with the $(j\text{ mod }n_i)$-th  and $[(j+1)\text{ mod }n_i]$-th qubits in the $i$-th layer into the same perceptron, where mod is the module operation. So that there are $n_{i+1}$ perceptrons between  $i$-th layer and $(i+1)$-th layer. The total number of parameters is 
	\begin{equation}
	\sum_{i=1}^{d-1}n_{i+1}\times N_p \approx O(N_{\text{All}}). 
	\end{equation}
	which means that the number of parameters grows linearly with the system size. This leads to the computational complexity $\mathcal{O}(N_{\text{All}}^3)$  to the DQFNN approach.
	%

\end{document}